\documentclass[12pt]{article}%
\usepackage{amsmath,latexsym}
\usepackage{graphicx}
\usepackage{amsmath}
\usepackage{amsfonts}
\usepackage{amssymb}%
\begin{document}

\title{{Conformal-symmetry wormholes supported by a perfect
     fluid}}
   \author{
Peter K F Kuhfittig*\\
\footnote{E-mail: kuhfitti@msoe.edu}
 \small Department of Mathematics, Milwaukee School of
Engineering,\\
\small Milwaukee, Wisconsin 53202-3109, USA}

\date{}
 \maketitle

\begin{abstract}\noindent
This paper presents a new wormhole solution by
assuming that a homogeneously distributed fluid
with equation of state $p=\omega\rho$ can be
adapted to an anisotropic spacetime such as a
wormhole and that this spacetime admits a
one-parameter group of conformal motions.  The
pressure $p$ in the equation of state becomes
the lateral pressure $p_t$ instead of the
radial pressure $p_r$, as assumed in previous
studies.  Given that $p_t=\omega\rho$, $p_r$
is then determined from the Einstein field
equations.  A wormhole solution can be
obtained only if $\omega <-1$ or $0<\omega <1$.
Since the former case corresponds to phantom
dark energy, which has been the subject of
earlier studies, we concentrate mainly on the
latter.  This case implies that given the
above conditions, dark matter can support
traversable wormholes.\\


\end{abstract}

\section{Introduction}\label{S:Introduction}

Wormholes are hypothetical handles or tunnels
in spacetime that connect different regions
of our Universe or completely different
universes altogether.  That wormholes could
be actual physical structures was first
proposed by Morris and Thorne \cite{MT88}.
These could be described by the static and
spherically symmetric line element
\begin{equation}\label{E:line1}
ds^{2}=-e^{2\Phi(r)}dt^{2}+\frac{dr^2}{1-b(r)/r}
+r^{2}(d\theta^{2}+\text{sin}^{2}\theta\,
d\phi^{2}),
\end{equation}
using units in which $c=G=1$.  Here
$\Phi=\Phi(r)$ is referred to as the
\emph{redshift function}, which must be
everywhere finite to avoid an event horizon.
The function $b=b(r)$ is called the
\emph{shape function} since it helps to
determine the spatial shape of the
wormhole when viewed, for example, in an
embedding diagram \cite{MT88}.  The
spherical surface $r=r_0$ is the radius of
the \emph{throat} of the wormhole.  Here
$b=b(r)$ must satisfy the following conditions:
$b(r_0)=r_0$, $b(r)<r$ for $r>r_0$, and
$b'(r_0)<1$, usually called the
\emph{flare-out condition}.  This condition
can only be satisfied by violating the null
energy condition (NEC), defined as follows:
for the stress-energy tensor $T_{\alpha\beta}$,
we must have
\begin{equation*}
  T_{\alpha\beta}k^{\alpha}k^{\beta}\ge 0
\end{equation*}
for all null vectors $k^{\alpha}$.  For
Morris-Thorne wormholes, matter that
violates the NEC is called ``exotic."

The equation of state (EoS) for a standard
perfect fluid, $p=\omega\rho$, $0<\omega <1$,
was studied a long time ago by Chandrasekhar
\cite{sC72}.  The realization that the
Universe is undergoing an accelerated
expansion \cite{aG01, sP99} has led to the
value of $\omega <-1/3$ due to the
Friedmann equation $\frac{\overset{..}{a}}{a}
=-\frac{4\pi}{3}(\rho+3p)$.  This case is
known as \emph{quintessence dark energy}.
The value of $\omega=-1$ corresponds to the
existence of Einstein's cosmological
constant \cite{mC01}.  The case that has
attracted the most attention in wormhole
physics is $\omega <-1$, referred to as
\emph{phantom dark energy} since this
case leads to a violation of the NEC:
given the null vector $(1,1,0,0)$,
$p+\rho =-\omega\rho +\rho <0$.  The EoS
$p=\omega\rho$, $0<\omega <1$, refers to
ordinary (baryonic) matter, as well as to
dark matter.

In this paper we also make use of conformal
symmetry, the existence of a conformal Killing
vector $\xi$ defined by the action of
$\mathcal{L_{\xi}}$ on the metric tensor
\begin{equation}
  \mathcal{L_{\xi}}g_{\mu\nu}=\psi(r)\,g_{\mu\nu};
\end{equation}
here $\mathcal{L_{\xi}}$ is the Lie derivative
operator and $\psi(r)$ is the conformal factor.

It is shown in this paper that the assumption of
conformal symmetry implies that a Morris-Thorne
wormhole is necessarily anisotropic.  It is then
assumed that a perfect-fluid distribution with
EoS $p=\omega\rho$ can be adapted to an
inhomogeneous spacetime such as a wormhole by
letting $p$ be the lateral pressure $p_t$
instead of $p_r$, as assumed in previous
studies \cite{fL05, sS05}.  It is subsequently
shown that a wormhole solution can exist only if
$\omega <-1$ or $0<\omega <1$.
%END OF SECTION

\section{Conformal Killing vectors and wormhole
    \\ construction}
This section consists of a brief discussion of the
assumption that our spacetime admits a one-parameter
group of conformal motions.  First we need to recall
that these are motions along which the metric tensor
of the spacetime remains invariant up to a scale
factor, which is equivalent to stating that there
exists a set of conformal Killing vectors such that
\begin{equation}\label{E:Lie}
   \mathcal{L_{\xi}}g_{\mu\nu}=g_{\eta\nu}\,\xi^{\eta}
   _{\phantom{A};\mu}+g_{\mu\eta}\,\xi^{\eta}_{\phantom{A};
   \nu}=\psi(r)\,g_{\mu\nu},
\end{equation}
where the left-hand side is the Lie derivative of the
metric tensor and $\psi(r)$ is the conformal factor.
Eq. (\ref{E:Lie}) shows that the vector $\xi$
characterizes the conformal symmetry since the metric
tensor $g_{\mu\nu}$ is conformally mapped into itself
along $\xi$.  It must be emphasized that the
assumption of conformal symmetry has proved to be
fruitful in numerous ways, not only leading to new
solutions but to new geometric and kinematical
insights \cite{HPa, HPb, MS93, sR08, fR10, fR12}.
Another fairly recent discovery is that the Kerr
black hole is conformally symmetric \cite{CMS10}.

Exact solutions of traversable wormholes admitting
conformal motions are discussed in Ref.
\cite{RRKKI} by assuming a noncommutative-geometry
background.  Two earlier studies assumed a
non-static conformal symmetry \cite{BHL07, BHL08}.

It is shown in Ref. \cite{pK15a} that the line element
\begin{equation}\label{E:line2}
   ds^2=- e^{\nu(r)} dt^2+e^{\lambda(r)} dr^2
   +r^2( d\theta^2+\text{sin}^2\theta\, d\phi^2)
\end{equation}
is particularly convenient for discussing the
consequences of the conformal-symmetry assumption.
In particular,
\begin{equation}\label{E:gtt}
   e^{\nu}=Cr^2
\end{equation}
and
\begin{equation}\label{E:grr}
  e^{\lambda}=\psi^{-2}.
\end{equation}
Moreover, the Einstein field equations are
\begin{equation}\label{E:E1}
\frac{1}{r^2}(1 - \psi^2)
  -\frac{(\psi^2)^\prime}{r}= 8\pi \rho,
\end{equation}
\begin{equation}\label{E:E2}
\frac{1}{r^2}(3\psi^2-1)= 8\pi p_r,
\end{equation}
and
\begin{equation}\label{E:E3}
   \frac{\psi^2}{r^2}+\frac{(\psi^2)'}{r}
      =8\pi p_t.
\end{equation}

It is clear from Eq. (\ref{E:gtt}) that the
wormhole spacetime cannot be asymptotically
flat.  So the wormhole material must be cut
off at some $r=a$ and joined to an exterior
Schwarzschild solution
 \begin{equation}
ds^{2}=-\left(1-\frac{2M}{r}\right)dt^{2}
+\frac{dr^2}{1-2M/r}
+r^{2}(d\theta^{2}+\text{sin}^{2}\theta\,
d\phi^{2}).
\end{equation}
We see from line element (\ref{E:line1}) that
$M=\frac{1}{2}b(a)$.  So for $e^{\nu}=Ca^2$,
we have $Ca^2=1-2M/a$ and the constant of
integration becomes
\begin{equation}
   C=\frac{1}{a^2}\left(1-\frac{b(a)}{a}
      \right).
\end{equation}
%END OF SECTION

\section{Wormhole solution}
In this section we take a closer look at the
EoS $p=\omega\rho$ as it relates to wormholes.
In a cosmological setting, we are dealing with a
homogeneously distributed fluid.  On the other
hand, for a given wormhole, the pressure may or
may not be isotropic, but a wormhole admitting
conformal motion is definitely not: suppose,
on the contrary, that $p_r=p_t$.  Then from
Eqs. (\ref{E:E2}) and (\ref{E:E3}),
\begin{equation}
   \frac{1}{r^2}(3\psi^2-1)=
  \frac{\psi^2}{r^2}+\frac{(\psi^2)'}{r}.
\end{equation}
After simplifying, we obtain the differential
equation
\begin{equation*}
   (\psi^2)'-\frac{2}{r}\psi^2=-\frac{1}{r},
\end{equation*}
which is linear in $\psi^2$ and readily solved
to obtain
\begin{equation*}
   \psi^2=\frac{1}{2}+cr^2.
\end{equation*}
We will see a bit later that to obtain a
wormhole solution, the equation must satisfy
the initial condition $\psi^2(r_0)=0$, where
$r=r_0$ is the throat of the wormhole.  The
result is
\begin{equation}
   \psi^2(r)=\frac{1}{2}-\frac{1}{2r_0^2}r^2.
\end{equation}
Now observe that for $r>r_0$, $\psi^2(r_0)<0$,
which is impossible since $\psi(r)$ is a
real-valued function.

Returning now to the EoS $p=\omega\rho$
describing a homogeneous distribution, it
is emphasized in Refs. \cite{fL05, sS05}
that an extension to an inhomogeneous
spherically symmetric spacetime is possible
by making $p$ the radial pressure $p_r$,
so that the transverse pressure  $p_t$ can
then be determined from the Einstein field
equations.

In this paper we follow the same strategy,
but instead of assuming that $p_r=\omega\rho$,
it is more convenient to use the EoS
\begin{equation}\label{E:EoS}
   p_t=\omega\rho;
\end{equation}
($p_r$ is then determined from the Einstein field
equations.)  Substituting in this equation,
\begin{equation}
   \frac{1}{8\pi}\left(\frac{\psi^2}{r^2}+
   \frac{(\psi^2)'}{r}\right)=\frac{\omega}{8\pi}
   \left[\frac{1}{r^2}(1-\psi^2)-
   \frac{(\psi^2)'}{r}\right].
\end{equation}
After simplifying, we obtain the differential
equation
\begin{equation}
   (\psi^2)'+\frac{1}{r}\psi^2=
        \frac{\omega}{1+\omega}\frac{1}{r},
\end{equation}
again linear in $\psi^2$.  The solution is
\begin{equation}
   \psi^2(r)=\frac{\omega}{1+\omega}+\frac{c}{r}.
\end{equation}
Comparing Eqs. (\ref{E:line1}) and
(\ref{E:line2}), observe that
\begin{equation}
   b(r)=r(1-e^{-\lambda(r)})=r[1-\psi^2(r)]
\end{equation}
by Eq. (\ref{E:grr}).  The condition
$b(r_0)=r_0$ at the throat implies that
$\psi^2(r_0)=0$, as noted earlier.  It follows
that the shape function is
\begin{equation}
   b(r)=r\left(1-\frac{\omega}{1+\omega}
   +\frac{1}{r}\frac{r_0\omega}{1+\omega}\right).
\end{equation}
To check the flare-out condition, we need to
find $b'(r_0)$:
\begin{equation}
   \left. b'(r_0)=1-\frac{\omega}{1+\omega}+
   \frac{1}{r}\frac{r_0\omega}{1+\omega}+
   r\left(-\frac{r_0\omega}{r^2(1+\omega)}
   \right)\right|_{r=r_0}=
   1-\frac{\omega}{1+\omega}.
\end{equation}
Now observe that $b'(r_0)<1$ only if
\begin{equation*}
   -\frac{\omega}{1+\omega}<0,\quad
        \omega\neq -1.
\end{equation*}
We conclude that $\omega <-1$ or $\omega >0$,
or, more completely,
\begin{equation}\label{E:omega}
   \omega <-1\quad \text{or}\quad 0<\omega <1.
\end{equation}

According to Ref. \cite{MT88}, the flare-out
condition is equivalent to $p_r+\rho <0$, i.e.,
the NEC is violated for the null vector
$(1,1,0,0)$.  Recalling the EoS $p_t=\omega\rho$,
Eq. (\ref{E:omega}) with $\omega <-1$ implies
that the NEC is violated for any null vector
of the form
\begin{equation*}
   (1,0,a,1-a),\quad 0\le a\le 1.
\end{equation*}
For $0<\omega <1$, the NEC is actually met.
It is readily checked, however, that
\begin{equation*}
   \left. p_r+\rho\right|_{r=r_0}=
   -\frac{1}{8\pi}\frac{1}{r_0^2}
   \frac{\omega}{1+\omega}<0
\end{equation*}
whenever
\[
   \omega <-1\quad\text{or}\quad 0<\omega <1,
\]
as before.

It is interesting to note that
\[
   p_r(r_0)=-\frac{1}{8\pi r_0^2},
\]
which is independent of $\omega$ and coincides
with $p_r(r_0)$ in Ref. \cite{MT88}.
%END OF SECTION

\section{The shadow universe}

Even though our starting point was the lateral
pressure in the EoS $p_t=\omega\rho$, it soon
became apparent that $p_r+\rho <0$ whenever
$\omega <-1$ or $0<\omega <1$.  In a cosmological
setting, we normally associate $\omega <-1$
with phantom dark energy and $0<\omega <1$ with
dark matter or normal (baryonic) matter.  It is
well known that phantom energy can support
traversable wormholes since the NEC is violated
\cite{fL05, sS05}.  So in the present situation,
the case $0<\omega <1$ is by far the more
interesting.

For the EoS $p_t=\omega\rho$, the conditions
$\omega <-1$ and $0<\omega <1$ cannot be
met simultaneously.  Well outside the
galactic halo, however, the case $\omega <-1$
would apply since on large scales, the Universe
is undergoing an accelerated expansion.  Inside
the galactic halo, on the other hand, dark
matter dominates since the galaxies themselves
do not participate in the expansion, being bound
gravitationally by predominantly dark matter.
So in the halo region, conformal-symmetry
wormholes can be supported by dark matter.
These wormholes would be part of what is often
called the shadow universe, normally invisible
to us.  Detection of wormholes may nevertheless
be possible by means of gravitational lensing
\cite{VE00, pK14, pK15b}.

For completeness let us note that it is shown
in Ref. \cite{RKRI} that the Navarro-Frenk-White
model can be used to show that dark matter can
support traversable wormholes provided that we
confine ourselves to the outer regions of the
galactic halo.  This restriction does not apply
to the present study.
%END OF SECTION

\section{Conclusion}

This paper discusses a new wormhole solution
by making the common and presumably
reasonable assumptions that a homogeneously
distributed cosmic fluid can be adapted to
an anisotropic spacetime such as a wormhole
and that this spacetime admits a one-parameter
group of conformal motions.  The pressure $p$
in the perfect-fluid EoS $p=\omega\rho$
becomes the lateral pressure $p_t$ instead
of the radial pressure $p_r$, as assumed in
previous studies.  Since $p_t=\omega\rho$ in
this paper, $p_r$ is determined from the
Einstein field equations.

After showing that the assumption of
conformal symmetry implies that any
Morris-Thorne wormhole must be anisotropic,
the EoS $p_t=\omega\rho$ is used to
determine the conformal factor $\psi(r)$
and hence the shape function $b=b(r)$.
Neither could exist unless $\omega <-1$
or $0<\omega <1$.  In both cases,
$p_r+\rho <0$, so that the NEC is
violated.

Well  outside the galactic halo region
the EoS $p=\omega\rho$, $\omega <-1$,
is normally interpreted as phantom dark
energy, which is known to support
traversable wormholes.  So the more
interesting case is $0<\omega <1$,
representing primarily dark matter in
the galactic halo region.  This shows
that given the above conditions, dark
matter can support traversable wormholes.

\end{document}